\documentclass[10pt]{article}
\usepackage[usenames,dvipsnames]{pstricks}
\usepackage{epsfig}
\usepackage{epstopdf}
\usepackage{amsmath}
\usepackage{slashed}
\usepackage{amssymb}
\usepackage{caption}
\usepackage{cite}
\usepackage{subcaption}
\usepackage{amsfonts}
\renewcommand{\today}{}
\begin{document}
\title{Non-Standard Interactions and Prospects for Studying Standard Parameter Degeneracies in DUNE and T2HKK}
\author {Surender Verma\thanks{Electronic address: s\_7verma@yahoo.co.in}  and Shankita Bhardwaj\thanks{Electronic address: shankita.bhardwaj982@gmail.com}}

\date{\textit{Department of Physics and Astronomical Science,\\Central University of Himachal Pradesh, Dharamshala 176215, INDIA.}}
\maketitle

\date{\today}
\begin{abstract}
The future long baseline experiments such as DUNE and T2HKK have promising prospects to determine the neutrino mass hierarchy and measuring standard $CP$ phase $\delta$. However, presence of possible non-standard interactions of neutrinos with matter may intricate this picture and is the subject matter of the present work. We have studied the standard parameter degeneracies in presence of non-standard interactions(NSI) with DUNE and T2HKK experiments. We examine the mass hierarchy degeneracy assuming (i) all NSI parameters to be non-zero and (ii) one NSI parameter($\epsilon_{e\mu}$) and its corresponding $CP$
phase($\delta_{e\mu}$) to be non-zero. We find that the later case is more appropriate to resolve mass hierarchy degeneracy with DUNE and T2HKK
experiments due to relatively small uncertainties emanating from the NSI sector. We have, also, investigated the octant degeneracy with
neutrino($\nu_{\mu}\rightarrow\nu_{e}$) and antineutrino($\bar{\nu}_{\mu}\rightarrow\bar{\nu}_{e}$) mode separately. We find that to resolve 
this degeneracy the long baseline experiment with combination of neutrino and antineutrino mode is essential.
Furthermore, we have considered DUNE in conjunction with T2HKK experiment to study $CP$ phase degeneracy due to standard($\delta$)
and non-standard($\delta_{e\mu}$) $CP$ phases. We find that DUNE and T2HKK, in conjunction, has more sensitivity for $CP$ violation effects(10$\sigma$ for true NH and 8.2$\sigma$ for true IH).  
\end{abstract}
\maketitle

\section{Introduction}
The discovery of non-zero neutrino masses and lepton flavor mixing by the reactor\cite{reac}, accelerator\cite{acc}, atmospheric\cite{atm}
and solar\cite{sol} neutrino oscillation experiments have revealed the values of oscillation parameters such as mass squared differences
$\Delta m_{21}^{2},|\Delta m_{31}^{2}|$ and mixing angles $\theta_{12},\theta_{23},\theta_{13}$\cite{theta13}, to an unprecedented accuracy.
At present, there are some unknown quantities in standard three-neutrino framework namely, $(i)$ sign of $\Delta m_{23}^{2}$, $(ii)$ the octant of
$\theta_{23}$ and $(iii)$ the $CP$-phase $\delta$, the determination of which is the prime objective of current and future neutrino oscillation
experiments. The difficulty in the determination of these unknowns is the existence of degeneracies in neutrino oscillation parameters.
To overcome these degeneracies, one of the method is to combine data from different neutrino oscillation experiments. Recently, this procedure
has been adopted by various studies\cite{pd1,pd2,pd3}, where the synergy between current
and future experiments has been considered. In principle, future neutrino oscillation experiments have sensitivity reach to perform precision
test of standard neutrino oscillation paradigm and to probe new physics beyond standard model(SM). In neutrino oscillation
experiments, one  model-independent way to study new physics(NP) is given by the framework of non-standard interactions(NSI)\cite{ohlsson,miranda}.

An alternative phenomena to explain neutrino flavor transitions, on the basis of NSI, was first proposed by Wolfenstein\cite{wolfenstein}.
Although, we know that they will show their effect in neutrino oscillation experiments at sub-leading level but, are important,
with the emergence of next generation experiments like Tokai-to-Hyper-Kamiokande(T2HK)\cite{t2hk}, Deep Underground Neutrino Experiment(DUNE) \cite{dune}, Tokai-to Hyper-Kamiokande-and-Korea(T2HKK)\cite{t2hkk} etc., where such type
of interactions can be probed. In general, the NSIs may manifest itself in propagation of neutrino through matter and the processes
involved in its creation and detection. The possible manifestations of NSIs have been widely studied in the literature and bounds on NSI
parameters have been derived from various experiments\cite{biggio,ohlsson2,ohlsson,miranda}. Furthermore, the model-independent bounds on NSI
in production and detection regions are an order of magnitude stronger than the matter NSI\cite{biggio}. In this work, we focus on matter NSI
which can be defined by dimension-six four-fermion operators given by\cite{wolfenstein,op2}
\begin{equation}\label{lagrangian}
 \mathcal{L}_{NSI}=2\sqrt{2}G_{F}\epsilon_{\alpha\beta}^{\zeta \mathcal{K}}\left[ \bar\nu_{\alpha}\gamma^{\rho}P_{L}\nu_{\beta}\right]
 \left[\bar\zeta\gamma_{\rho}P_{\mathcal{K}}\zeta\right]+ h.c.,
\end{equation}
where $\alpha,\beta=e,\mu,\tau$, $\mathcal{K}=L,R$, $\zeta=u,d,e$ and $\epsilon_{\alpha\beta}^{\zeta \mathcal{K}}$ are dimensionless
parameters indicating the strength of the new interaction having units of $G_{F}$.
To probe matter NSI long baseline neutrino experiments(LBNE)
are ideal and the neutral current(NC) interactions which affect the neutrino propagation coherently can also be studied at far detectors.
The next generation LBNE such as DUNE, T2HK and T2HKK may reach the sensitivity to reveal NSI in neutrino sector.

In the leptonic sector, the $CP$ violation can render leptogenesis mechanism which in turn may shed light on baryogenesis\cite{lep1,lep2}.
It is very difficult to measure leptonic $CP$-violation in presence of NSI as it will get bewildered by the existence of possible $CP$-violation
generated by NSI itself. Undoubtedly, the existence of NSI have opened an entirely new window to explore NP beyond standard model.

Previously, the authors of ref.\cite{pd1} have explored the possibility to disentangle the $CP$ violating effects due to standard and non-standard
contributions under the assumption that only one NSI parameter $\epsilon_{e\mu}$ or $\epsilon_{e\tau}$ is present. In ref.\cite{pd2}, 
the parameter degeneracies in LBNE originating from non-standard interactions have been studied and ref.\cite{danny} has focused on NSI at DUNE, T2HK 
and T2HKK and has concluded that overall DUNE has the best sensitivity to the magnitude of the NSI parameters, while T2HKK has the best $CP$ violation
sensitivity with or without NSI. Furthermore, in ref.\cite{goswami1}, the authors have studied the impact of non-zero NSI on the $CP$ precision of DUNE.
The authors of refs.\cite{mehta1,mehta2} and \cite{goswami2} have explored the effects of NSI on $CP$ violation sensitivity and hierarchy
sensitivity at DUNE, respectively. In ref.\cite{pd3} the authors have studied the sensitivity to mass hierarchy,
the octant of $\theta_{23}$ and $CP$ phase $\delta$ in the future long baseline experiments T2HK and DUNE assuming standard interactions(SI) only.
In general, earlier studies on standard parameters degeneracies with SI or matter NSI have mostly focused on DUNE\cite{mehta1,mehta2,goswami2,sk}. Motivated by the long baseline of
T2HKK experiment, it is imperative to study physics potential of T2HKK and DUNE+T2HKK, in resolving standard parameter degeneracies in presence of NSI.
In the present work, we have investigated prospects for lifting mass hierarchy degeneracy(sign degeneracy), $\theta_{23}$-octant degeneracy and
$CP$-phase degeneracy in DUNE, T2HKK and DUNE+T2HKK with matter NSIs.

T2HKK is a long baseline experiment proposed to enhance the hierarchy sensitivity of T2HK by setting one of the two tanks of HK detector at a site
in Korea. This multi-detector set-up is advantageous as it give access to a longer baseline of 1100 km and simultaneously boost the data
at the T2HK with baseline of 295 km\cite{dunet2hkk}. The neutrino oscillation probabilities are strongly affected by the matter effects
in long baseline experiments. These matter effects can be beneficial in lifting up the standard parameter degeneracies. Therefore, we have 
considered the T2HKK setup with larger baseline of 1100 km in the analysis. In present work, we have studied the standard parameter degeneracies
i.e. mass hierarchy degeneracy and octant degeneracy in presence of matter NSI with DUNE and T2HKK experiment. We have assumed all NSI parameters
to be non-zero in one case and only one off-diagonal NSI parameter $\epsilon_{e\mu}$ to be non-zero, in another case. We find that the later case
is better at resolving standard parameter degeneracies in case of both DUNE and T2HKK experiments. Due to the larger baseline, T2HKK is found to
have similar sensitivity as DUNE experiment to resolve standard parameter degeneracies including NSI. Furthermore, we have investigated
the $CP$ phase degeneracy occurring due to the contribution from standard and non-standard $CP$ phases. We observe that it is difficult to disentangle the $CP$ effects due to SI phase $\delta$ from NSI phase $\delta_{e\mu}$ at DUNE+T2HKK experiment as this conjunction is more sensitive to study $CP$ violation effects\cite{dunet2hkk}.

We organize the paper as follows: in section II, we present the formalism to write oscillation probability in presence of matter NSIs. We discuss about
the long baseline experiments DUNE, T2HKK and corresponding simulation details in section III. In section IV, we discuss the prospects to resolve
standard parameter degeneracies in these LBNEs. We have presented our results and, subsequent, discussion in section V. Finally, we conclude in
section VI.

\section{Formalism: Oscillation Probabilities}
The Hamiltonian for the neutrino propagation in presence of matter NSI can be written as,
\begin{equation}~\label{hamiltonian}
 H=\frac{1}{2E}\left[U diag(0,\Delta m_{21}^{2},\Delta m_{31}^{2}){U}^{\dagger}+V\right],
\end{equation}
where, $U$ is the PMNS mixing matrix containing three mixing angles$(\theta_{ij},i<j=1,2,3)$ and one $CP$ phase $\delta$, 
$\Delta m_{ji}^{2}\equiv m_{j}^{2}-m_{i}^{2}$. $V$ is the matter potential due to interaction of neutrino with matter, viz.,
\begin{equation}~\label{matterpotential}
V=\mathcal{A}\left(
\begin{array}{ccc}
1+\epsilon_{ee} & \epsilon_{e\mu}e^{i\delta_{e\mu}} & \epsilon_{e\tau}e^{i\delta_{e\tau}} \\
\epsilon_{e\mu}e^{-i\delta_{e\mu}} & \epsilon_{\mu\mu} & \epsilon_{\mu\tau}e^{i\delta_{\mu\tau}}  \\
\epsilon_{e\tau}e^{-i\delta_{e\tau}}&\epsilon_{\mu\tau}e^{-i\delta_{\mu\tau}}&\epsilon_{\tau\tau}
\end{array}
\right),
\end{equation}
where, $\mathcal{A}\equiv 2\sqrt{2}G_{F}N_{e}(r)E$. The unit contribution in the first element of the matrix $V$ is due
the matter term contribution from standard charged-current interactions. The diagonal element of $V$ are real i.e, $\delta_{\alpha\beta}=0$ (where $\alpha,\beta=e,\mu,\tau$) 
for $\alpha=\beta$ and $\epsilon_{\alpha\beta}\equiv \sum_{\zeta,\mathcal{K}}\epsilon_{\alpha\beta}^{\zeta \mathcal{K}}\frac{N_{\zeta}}{N_{e}}$.
The oscillation probability for $\mu\rightarrow e$ channel can be written as\cite{pd2}

\begin{eqnarray}~\label{probability}
\nonumber
 P(\nu_{\mu}\rightarrow\nu_{e})= && p^{2}f^{2}+2pqfg\cos\left(\Delta+\delta\right)+q^{2}g^{2}\\
 \nonumber
                                 && +4\hat{A}\epsilon_{e\mu}\{pf[s_{23}^{2}f\cos\left(\delta_{e\mu}+\delta\right)
                                 +c_{23}^{2}g\cos\left(\Delta+\delta+\delta_{e\mu}\right)]\\ \nonumber
                                 && +qg[c_{23}^{2}g\cos\delta_{e\mu}+s_{23}^{2}f\cos(\Delta-\delta_{e\mu})]\}\\
                                 \nonumber
                                 &&+4 \hat{A}\epsilon_{e\tau}s_{23}c_{23}\{pf\left[f\cos(\delta_{e\tau}+\delta)-g\cos(\Delta+\delta+\delta_{e\tau})\right]\\
                                 \nonumber
                                 &&-qg\left[g\cos\delta_{e\tau}-f\cos(\Delta-\delta_{e\tau})\right]\}+4\hat{A}^2g^{2}c_{23}^{2}|c_{23}\epsilon_{e\mu}-s_{23}
                                 \epsilon_{e\tau}|^2\\
                                 \nonumber
                                 &&+4\hat{A}^2f^{2}s_{23}^2|s_{23}\epsilon_{e\mu}+c_{23}\epsilon_{e\tau}|^2\\
                                 \nonumber
                                 &&+8\hat{A}^2fgs_{23}c_{23}\{c_{23}\cos\Delta[s_{23}(\epsilon_{e\mu}^2-\epsilon_{e\tau}^{2})\\
                                 \nonumber
                                 &&+2c_{23}\epsilon_{e\mu}\epsilon_{e\tau}\cos(\delta_{e\mu}-\delta_{e\tau})]\\
                                 \nonumber
                                 &&-\epsilon_{e\mu}\epsilon_{e\tau}
                                 \cos(\Delta-\delta_{e\mu}+\delta_{e\tau})\}\\
                                 &&+\mathcal{O}(s_{13}^2\epsilon,s_{13}\epsilon^{2},\epsilon^{3}),
\end{eqnarray}
\begin{eqnarray}~\label{parameter}
p\equiv2s_{13}s_{23},q\equiv2rs_{12}c_{12}c_{23},r=|\Delta m_{21}^{2}/\Delta m_{31}^{2}|,(f,\bar{f})\equiv \frac{\sin[\Delta(1\mp\hat{A}(1+\epsilon_{ee}))]}{(1\mp\hat{A}(1+\epsilon_{ee}))},\\
g\equiv \frac{\sin(\hat{A}(1+\epsilon_{ee})\Delta)}{\hat{A}(1+\epsilon_{ee})}, \Delta\equiv\bigg|\frac{\Delta m_{31}^{2} L}{4E}\bigg|,
\hat{A}\equiv\bigg|\frac{A}{\Delta m_{31}^{2}}\bigg|.
\end{eqnarray}

where, $s_{ij}=\sin\theta_{ij},c_{ij}=\cos\theta_{ij},i<j$, $(i,j)=1,2,3$. Similar expression can be obtained for inverted hierarchy(IH) by replacing 
$\Delta m_{31}^{2}\rightarrow-\Delta m_{31}^{2}$ (i.e. $\Delta\rightarrow-\Delta, \hat{A}\rightarrow-\hat{A}, f\rightarrow-\bar{f}, g\rightarrow-g$ 
and $q\rightarrow-q$). The expression for antineutrino oscillation probability can be written by replacing $\hat{A}\rightarrow-\hat{A},
\delta\rightarrow-\delta$ and $\delta_{\alpha\beta}=-\delta_{\alpha\beta}$ in Eqn. (\ref{probability}).

\section{Experimental Setups}
Considering the sensitivity reach of the present and future long baseline neutrino oscillation experiments(for example, DUNE and T2HKK), it is very
important to study the individual and collective effects of NSI parameters on parameter degeneracies. We have used GLoBES package\cite{glb1, glb2}
with best-fit values and ranges of the standard neutrino mixing parameters, as given in \cite{data} to simulate the DUNE and T2HKK.
The current bounds on NSI parameters used in present analysis are $\epsilon_{ee}<4.2,|\epsilon_{e\mu}|<0.33,|\epsilon_{e\tau}|
<3.0,\epsilon_{\mu\mu}<0.068,|\epsilon_{\mu\tau}|<0.04,\epsilon_{\tau\tau}<0.15$ \cite{biggio}.  The $CP$ phases 
$\delta_{\alpha\beta}$ of the off-diagonal NSI parameters are still unconstrained and can lie in the range $
\delta_{\alpha\beta}\in[-\pi,+\pi]$.

\begin{table}[h]

  \centering
 \begin{tabular}{c c c}
 \hline
 \hline
{NSI Parameter} & True value & Marginalization range \\
  \hline
   $\epsilon_{ee}$ & 0.4 & [-0.4,0.4]\\ 
   $\epsilon_{\mu\mu}$ & 0.03  & [-0.05,0.05]\\
   $\epsilon_{\tau\tau}$ &  0.1 & [-0.15,0.15]\\ 
   $\epsilon_{e\mu}$ & 0.04 & [0,0.10]\\
   $\epsilon_{e\tau}$ &  0.04 & [0,0.10]\\ 
   $\epsilon_{\mu\tau}$ & 0.04 & [0,0.04]\\
   $\delta_{e\mu}$& [-$\pi,\pi$]&[-$\pi,\pi$]\\
   $\delta_{e\tau}$& [-$\pi,\pi$]&[-$\pi,\pi$]\\
   $\delta_{\mu\tau}$& [-$\pi,\pi$]&[-$\pi,\pi$]\\
   \hline
   \hline
   \end{tabular}
  \caption{\label{tab1} The true values and marginalization ranges for all NSI parameters used in the analysis\cite{biggio,pm1}.}
\end{table}

The experimental configurations, energy resolutions and systematic uncertainties considered in the present work are as follows:

\subsection{DUNE} The DUNE experiment\cite{dune}, situated in the USA, is a globally synchronized endeavor of neutrino physicists around the world.
Out of many others, the neutrino physics goals of the experiment are to unravel the sign of neutrino mass hierarchy($\Delta m_{31}^2$) and to measure
the $CP$ phase(s). The experiment is planned to direct neutrino beam from Fermilab to Homestake mine in South Dakota providing an optimum baseline of
1300 km for manifestation of matter effects in neutrino oscillations. Unlike Hyper-K, DUNE is an on-axis experiment. We have used DUNE
CDR\cite{dune,dunecdr} with 35 kt LAr far detector. The optimized beam design that employs 80 GeV beam of protons having 1.0 MW power have
been used to simulate the experiment. We have considered 5(+5) years of run in neutrino(antineutrino) mode resulting in an exposure of 350 kt.MW.years. The appearance efficiency ($\epsilon_{app}$), energy resolutions($E_{R_e},E_{R_\mu}$) taken in the present analysis is 80$\%$, (0.15$/\sqrt{E}$, 0.2$/\sqrt{E}$), respectively. The normalization and energy calibration uncertainty for $\nu_e$ signal($N_S,E_S$)/background($N_B,E_B$) is taken to be $N_S=5\%$, $E_S=2\%$, $N_B=10\%$ and $E_B=10\%$. For $\nu_\mu$ signal($N_S,E_S$)/background($N_B,E_B$) the values are $N_S=5\%$, $E_S=5\%$, $N_B=10\%$ and $E_B=10\%$.

\subsection{T2HKK} 

The T2HKK experiment\cite{t2hkk}, an extension T2HK\cite{t2hk}, is proposed to be stretched over Japan and Korea. The neutrino beam will be directed from J-PARC facility in Japan  to two water-Cherenkov detectors: (i) first detector at Kamioka mine in Japan with a baseline of 295 km (ii) second detector to be built in Korea  providing a baseline of 1100 km. In the present work, we have considered 1100 km baseline(also, referred as T2HKK), with detector at $1.5^{o}$ off-axis with respect the neutrino beam,  where matter effects will be large.  We choose
13 MW.years beam power which is similar to that of T2HK. The running time, in ratio 1:3 for neutrino and antineutrino mode, is 10 years amounting to total exposure of 2.7$\times 10^{22}$ protons on target(POT). The appearance efficiency ($\epsilon_{app}$), energy resolutions($E_{R_e},E_{R_\mu}$) taken in the present analysis is 50$\%$, (0.085$/\sqrt{E}$, 0.085$/\sqrt{E}$), respectively. The normalization and energy calibration uncertainty for $\nu_e$ signal($N_S,E_S$)/background($N_B,E_B$) is taken to be $N_S=5\%$, $E_S=0.01\%$, $N_B=5\%$ and $E_B=0.01\%$. For $\nu_\mu$ signal($N_S,E_S$)/background($N_B,E_B$) the values are $N_S=2.5\%$, $E_S=0.01\%$, $N_B=20\%$ and $E_B=0.01\%$.
 
 \section{Parameter Degeneracies}
\begin{figure}[htp]
\centering 
 \epsfig{file=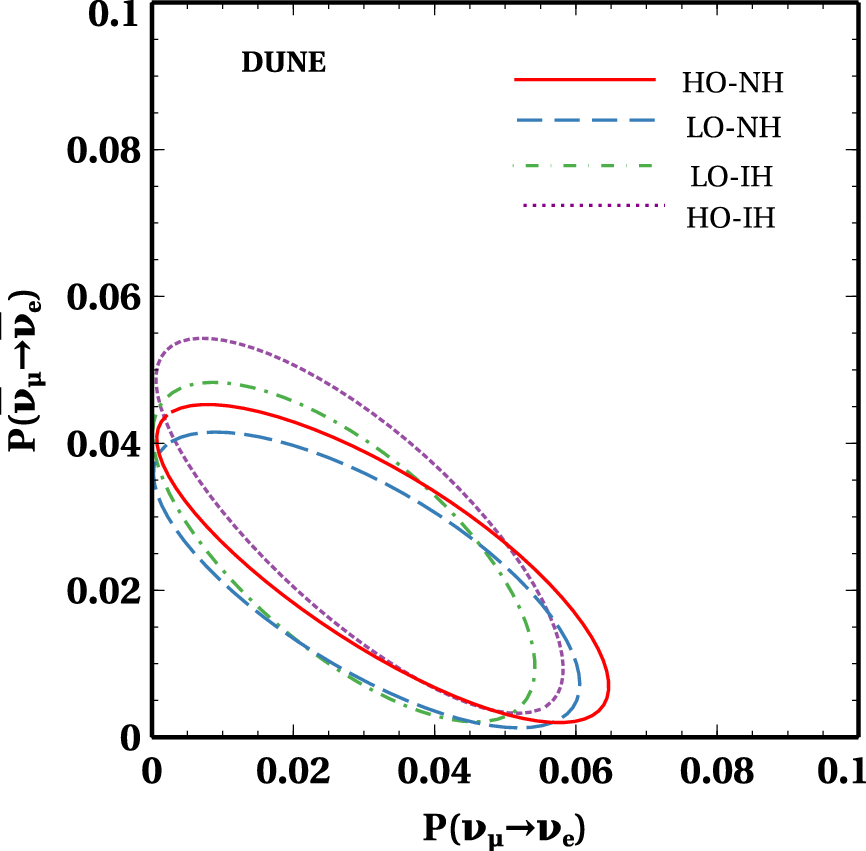, height=6cm, width=7cm}
 \caption{~\label{fig0} The bi-probability plots with $\epsilon_{e\mu}=0.04$ and $CP$-phase $\delta_{e\mu}=\pi/2$ for DUNE with baseline $L=1300$ km and $E=3$ GeV. The standard $CP$-phase $\delta$ is varied from $0$ to $2\pi$.}
\end{figure}

In general, different set of oscillation parameters may give identical predictions for oscillation probability resulting in parameter
degeneracies and making it difficult to uniquely determine these parameters. The bi-probability plots are well known constructions to study
parameter degeneracies in presence of NSI parameters and $CP$ phases. In presence of off-diagonal NSI the neutrino/antineutrino
oscillation probability exhibit degeneracy between SI and NSI phase when $P^{SI}(\delta)=P^{NSI}(\phi,\epsilon_{e\mu},\delta_{e\mu})$ and $\bar{P}^{SI}(\delta)=\bar{P}^{NSI}(\phi,\epsilon_{e\mu},\delta_{e\mu})$,
where $\phi$ is Dirac $CP$ phase in a model with NSI. If we presume that mixing angles and mass squared differences are known from some other experiment, then for every value of SI phase $\delta$ there will be three unknowns($\phi$, $\epsilon_{e\mu}$ and the phase $\delta_{e\mu}$) which generate an off-diagonal NSI degeneracy. Accordingly, a measurement of $P$
and $\bar{P}$ in an experiment provide two constraints, for each value of $\delta$, a solution for $\epsilon_{e\mu}$ and $\delta_{e\mu}$ will exist for any value of $\phi$ resulting in parameter degeneracy. As a representative plot to depict standard parameter degeneracies we have shown, in Fig.(\ref{fig0}), mass hierarchy and octant degeneracy assuming NSI parameters $\epsilon_{e\mu}=0.04,\delta_{e\mu}=\pi/2$ and varying $\delta$ from $0$ to $2\pi$ for DUNE experiment. All other NSI parameters are assumed to be zero. The solid and dashed ellipses correspond to higher octant(HO) and lower octant(LO) of $\theta_{23}$, respectively, for normal hierarchy(NH). Similarly, the dotted and dash-dotted ellipses  correspond to higher octant(HO) and lower octant(LO) of $\theta_{23}$, respectively, for inverted hierarchy(IH). The significant overlapping of the ellipses show four-fold $\theta_{23}$ octant and mass hierarchy degeneracy. For example, points of intersection of solid(dotted) and dashed(dash-dotted) ellipses exhibit $\theta_{23}$ octant degeneracy as neutrino and antineutrino oscillation probabilities are same in both cases for normal(inverted) hierarchy. 
 
\subsection{Mass hierarchy degeneracy}
\begin{figure}[htp]
\centering
\epsfig{file=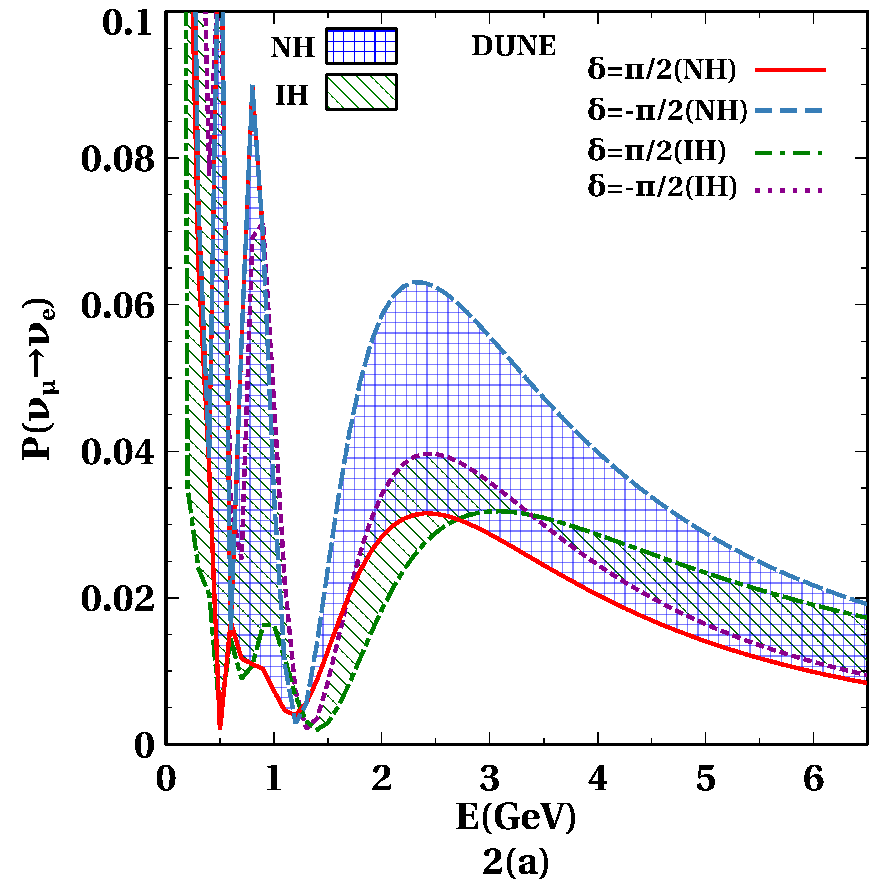, height=6cm, width=6cm}\hspace{1cm}
\epsfig{file=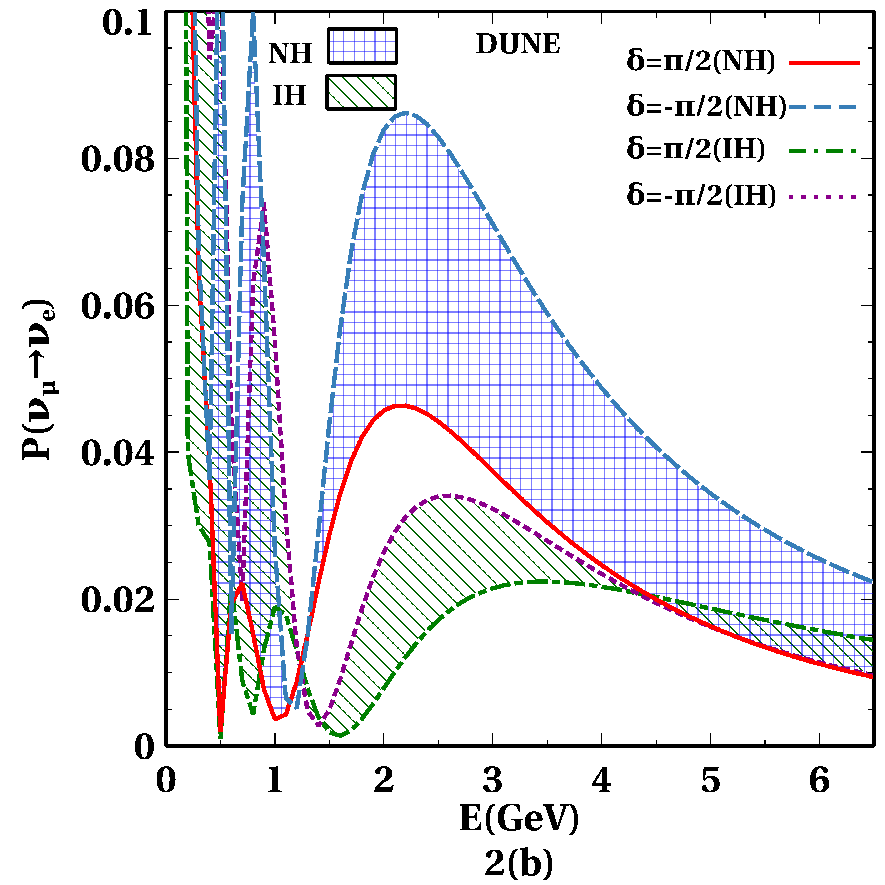, height=6cm, width=6cm}
\epsfig{file=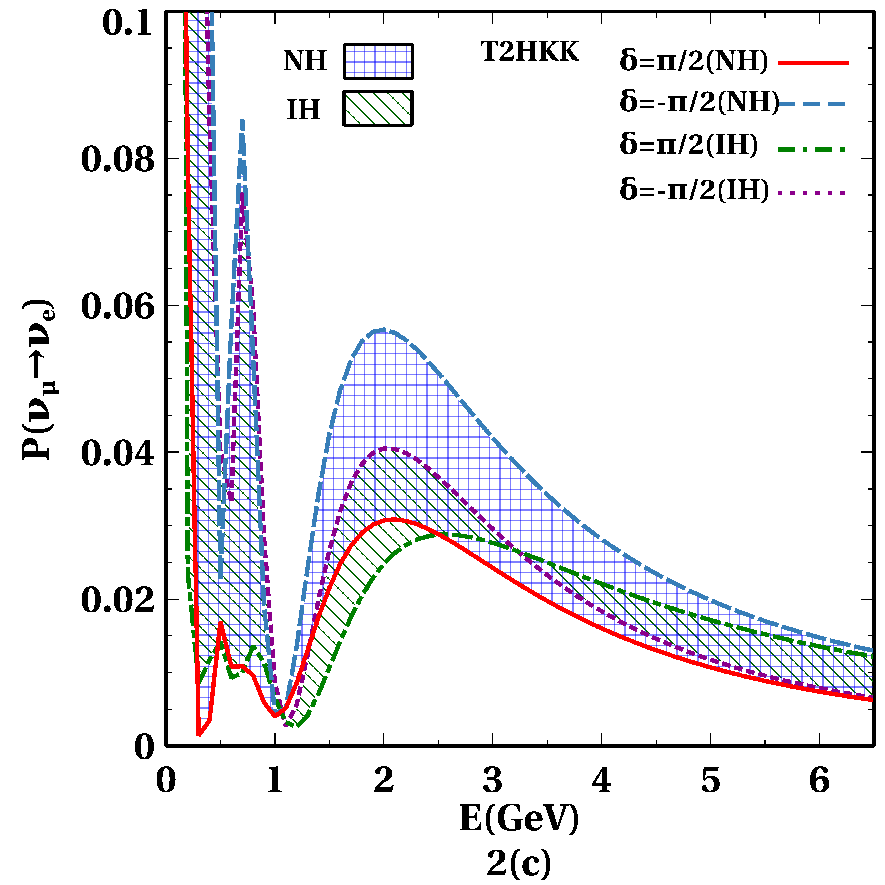, height=6cm, width=6cm}\hspace{1cm}
\epsfig{file=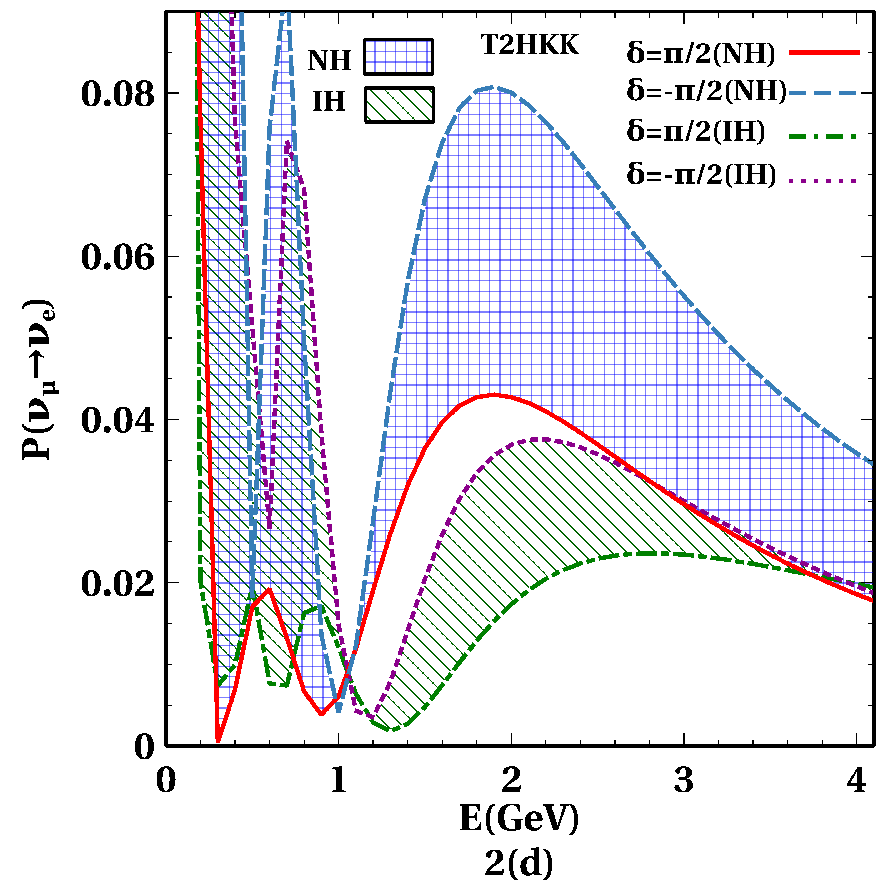, height=6cm, width=6cm}
\caption{~\label{fig1} The appearance probability $P(\nu_{\mu}\rightarrow\nu_{e})$ as function of neutrino beam energy $E$ for DUNE(first row)
and T2HKK(second row). The left(right) panel describes mass hierarchy assuming all($\epsilon_{e\mu}, \delta_{e\mu}$) NSI parameters non-zero. $\theta_{23}=42^{o}$($\theta_{23}=48^{o}$) for left panel(right panel). The band comes due to variation of $\delta,\delta_{e\mu}\in [-\pi,\pi]$ and boundaries correspond to $\delta=\pm\pi/2$
for NH as well as IH.}
\end{figure}

In the determination of unknown neutrino mixing parameters an ambiguity exists in correlated way between $\delta$ and  sign of $\Delta m_{31}^{2}$.
The sign of $\Delta m_{31}^{2}$ can be determined by measuring interference between the vacuum and the matter effects. The simultaneous determination
of $\delta$ and sign of $\Delta m_{31}^{2}$ can be done in long baseline experiments in both standard\cite {pd3} and non-standard cases \cite{pd1}.
Thus, we study the prospects to resolve this degeneracy with two future long baseline experiments DUNE and T2HKK involving NSIs.
Throughout this work we have considered the $\nu_\mu\rightarrow\nu_e$ channel to study various parameter degeneracies.
We have obtained plots for DUNE and T2HKK for two cases assuming (i) all NSI parameters non-zero i.e. $\epsilon_{ee}=0.4,\epsilon_{e\mu}=0.04
,\epsilon_{e\tau}=0.04,\epsilon_{\mu\mu}=0.03,\epsilon_{\mu\tau}=0.04,\epsilon_{\tau\tau}=0.1$ and NSI $CP$ phases $\delta_{\alpha\beta}=[-\pi,\pi]$
(ii) only $\epsilon_{e\mu},\delta_{e\mu}$ are non-zero ($\epsilon_{e\mu}=0.04, \delta_{e\mu}=[-\pi,\pi]$)(Fig.(\ref{fig1})). 

\begin{figure}[htp]
\centering
\epsfig{file=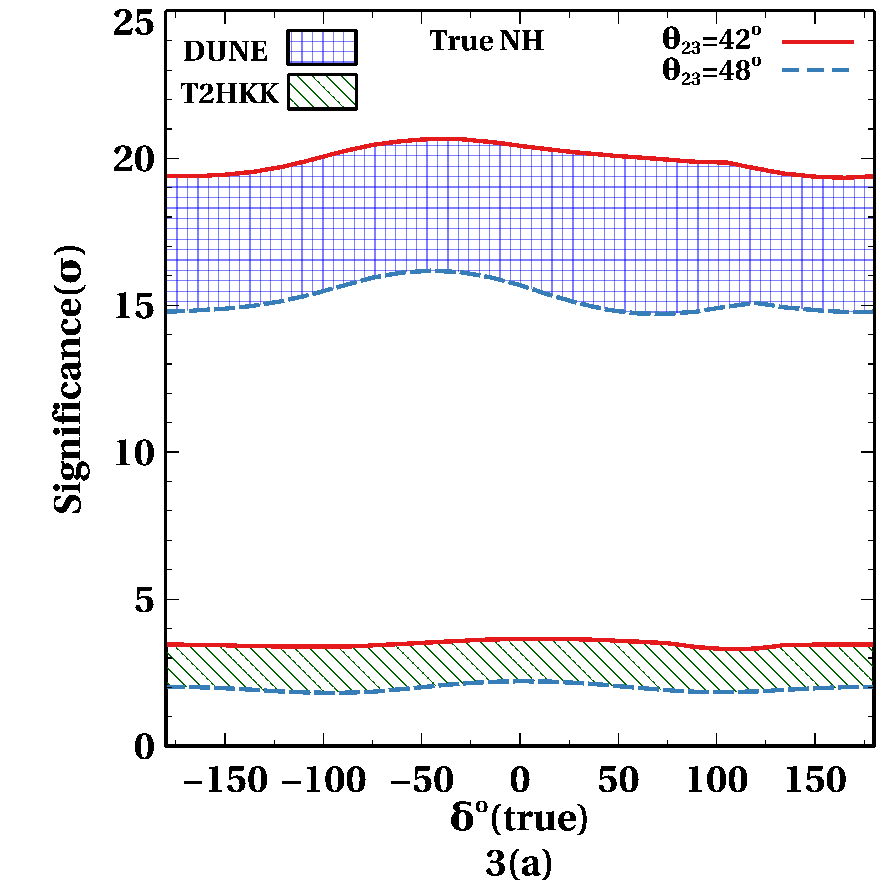, height=6cm, width=6cm}\hspace{1cm}
\epsfig{file=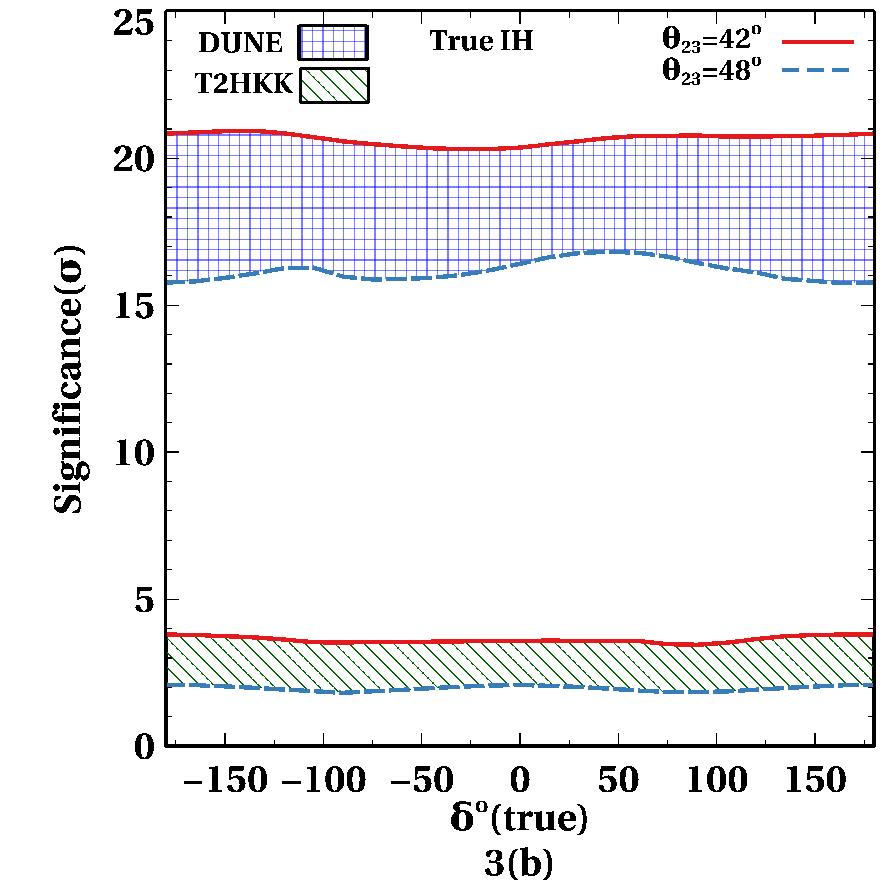, height=6cm, width=6cm}
\epsfig{file=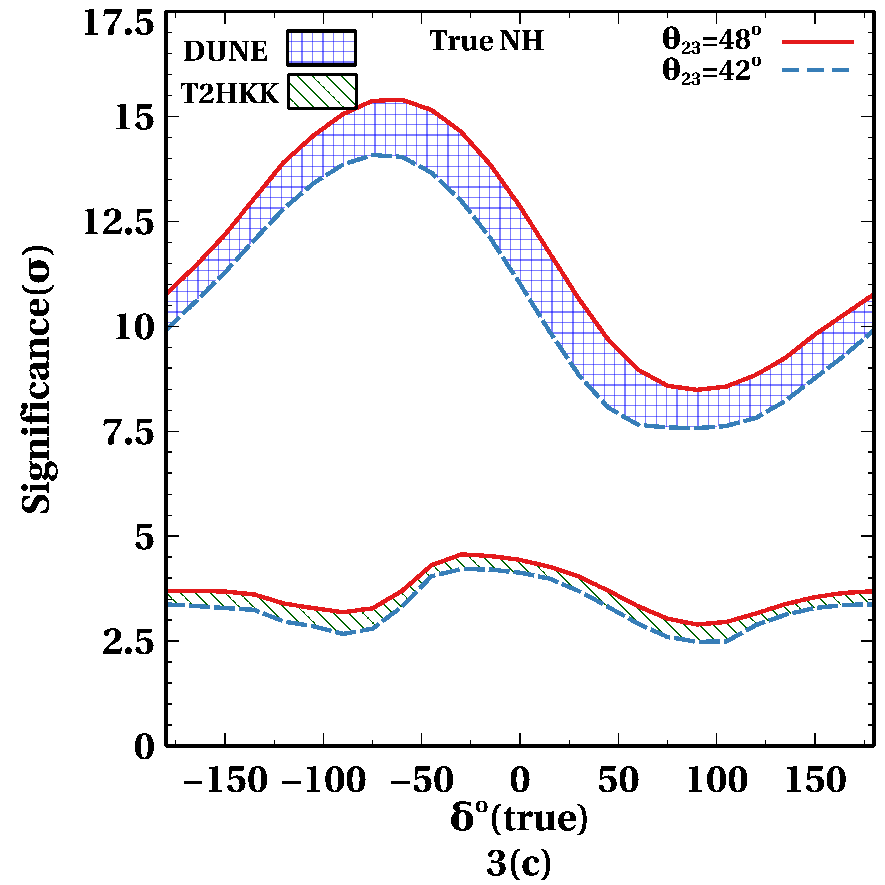, height=6cm, width=6cm}\hspace{1cm}
\epsfig{file=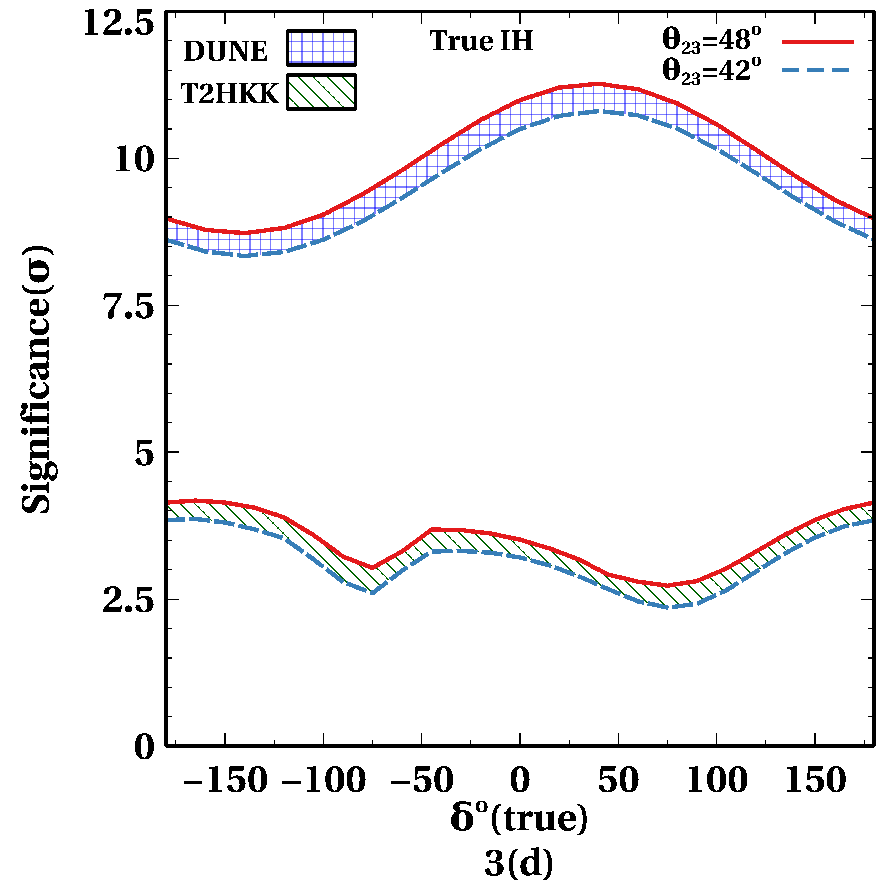, height=6cm, width=6cm}
\epsfig{file=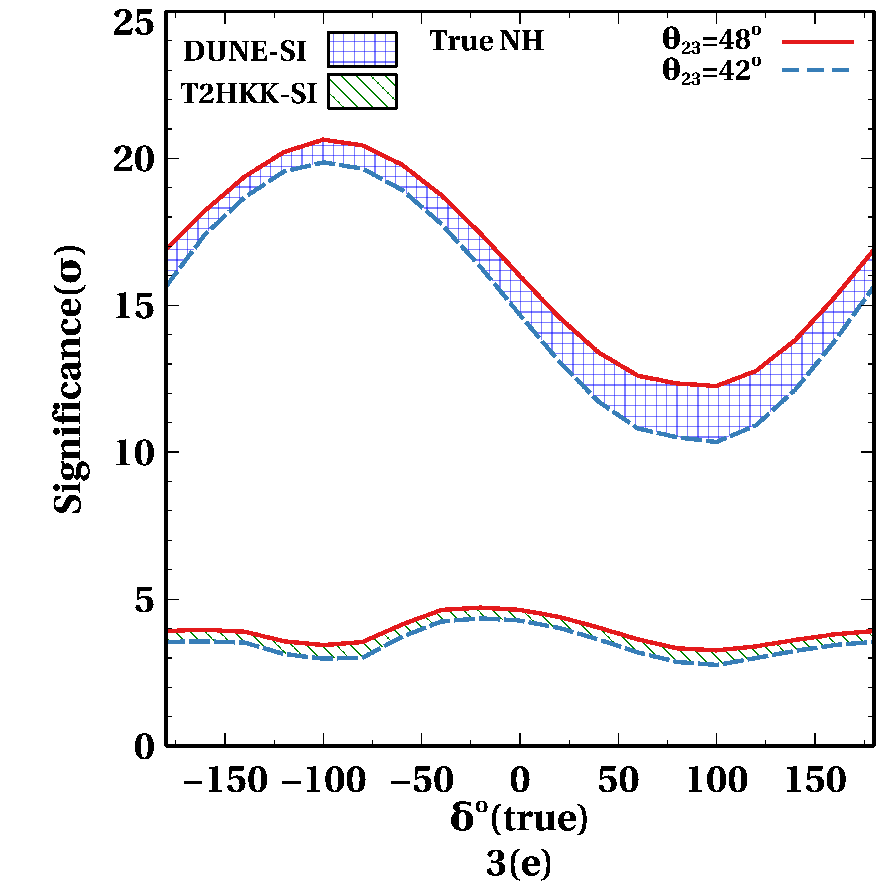, height=6cm, width=6cm}\hspace{1cm}
\epsfig{file=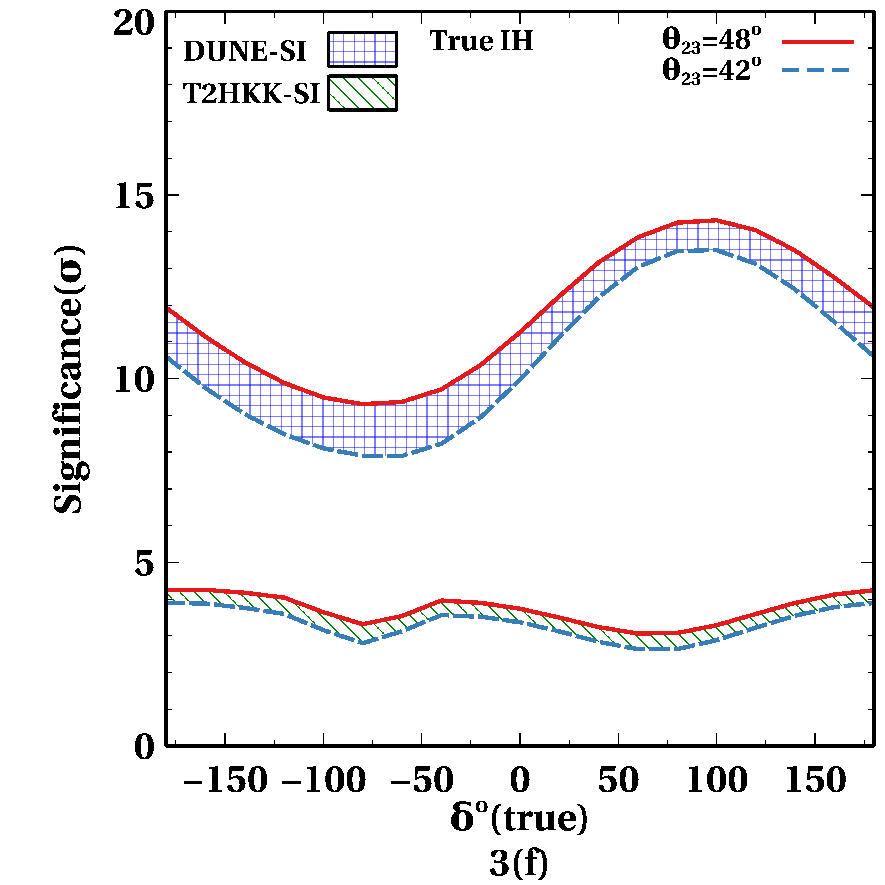, height=6cm, width=6cm}
\caption{~\label{fig1a}Sensitivity plots for mass hierarchy. The plots for significance($\sqrt{\chi^{2}}$) as a
function of $\delta(true)$ with true NH(left panel) and true IH(right panel). The first(second) row describes mass hierarchy sensitivity assuming
all($\epsilon_{e\mu}, \delta_{e\mu}$) NSI parameters non-zero. Figs. 3(e) and 3(f) show the sensitivity plots for SI case. The band comes due to variation of $\theta_{23}$
and boundaries correspond to $\theta_{23}=42^{o}$ and $48^{o}$.}
\end{figure}

We have, also, shown sensitivity plots for mass hierarchy for cases (i) and (ii) in Figs.(\ref{fig1a}(a))-(\ref{fig1a}(b)) and 
Figs.(\ref{fig1a}(c))-(\ref{fig1a}(d)), respectively. We plot significance $(\sigma=\sqrt{\chi^{2}})$  as a function of $\delta(true)$ to study hierarchy
sensitivity of DUNE and T2HKK experiments for true NH(true IH) in left panel(right panel) of Fig.(\ref{fig1a}). The true NH(true IH) sensitivity
plot is obtained by considering NH(IH) in the true spectrum and IH(NH) in test spectrum. We have marginalized over $\delta,\theta_{23}$
and $\epsilon$ by considering them in test spectrum. The statistical definition of $\chi^{2}$ for understanding the aspects of the mass hierarchy sensitivity plot in case of true NH is as follows:

\begin{eqnarray}
\nonumber
 \chi^{2}_{NH}\equiv &&\min_{(\delta,\theta_{23},\epsilon)_{test}}\\ \nonumber
 &&\sum_{i=1}^{x}\sum_{j,k=1}^{2}\frac{\bigl[N_{NH}^{i,j,k}(\delta,\theta_{23},\epsilon)_{true}-N_{IH}^{i,j,k}(\delta,\theta_{23},
                \epsilon)_{test}\bigr]^{2}}
              {N_{NH}^{i,j,k}(\delta,\theta_{23},\epsilon)_{true}},\\
\end{eqnarray}
where, $N_{NH}^{i,j,k}$ and $N_{IH}^{i,j,k}$ denote the number of true events and test events for NH and IH in the $(i,j,k)^{th}$ bin, respectively. The index $i$ runs over 1 to $x$, $x$ is the number of bins for particular experiment. For DUNE, $x=39$, of 250 MeV width in 0.5-10 GeV range and for T2HKK $x=20$, of 40 MeV width in 0.4-1.2 GeV range. The index $j$ describe type of mode
i.e. neutrino or antineutrino. $j=1(j=2)$ for neutrino(antineutrino) mode. The index $k$ describe type of channel considered i.e. appearance or disappearance. $k=1(k=2)$ for appearance(disappearance) channel. The NSI parameter
$\epsilon=\epsilon_{\alpha\beta}e^{i\delta_{\alpha\beta}}$. We have minimized over the marginalization ranges of NSI parameters
as given in Table \ref{tab1}.

\subsection{Octant degeneracy}
\begin{figure}[htp]
\centering
\epsfig{file=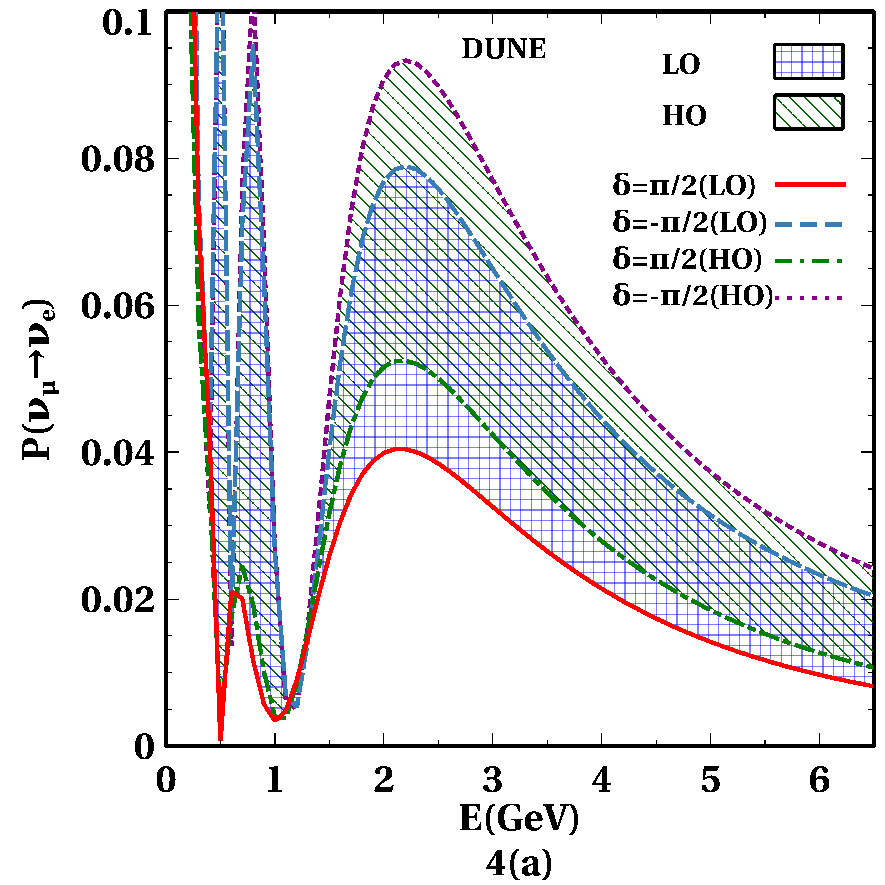, height=6cm, width=6cm}\hspace{1cm}
\epsfig{file=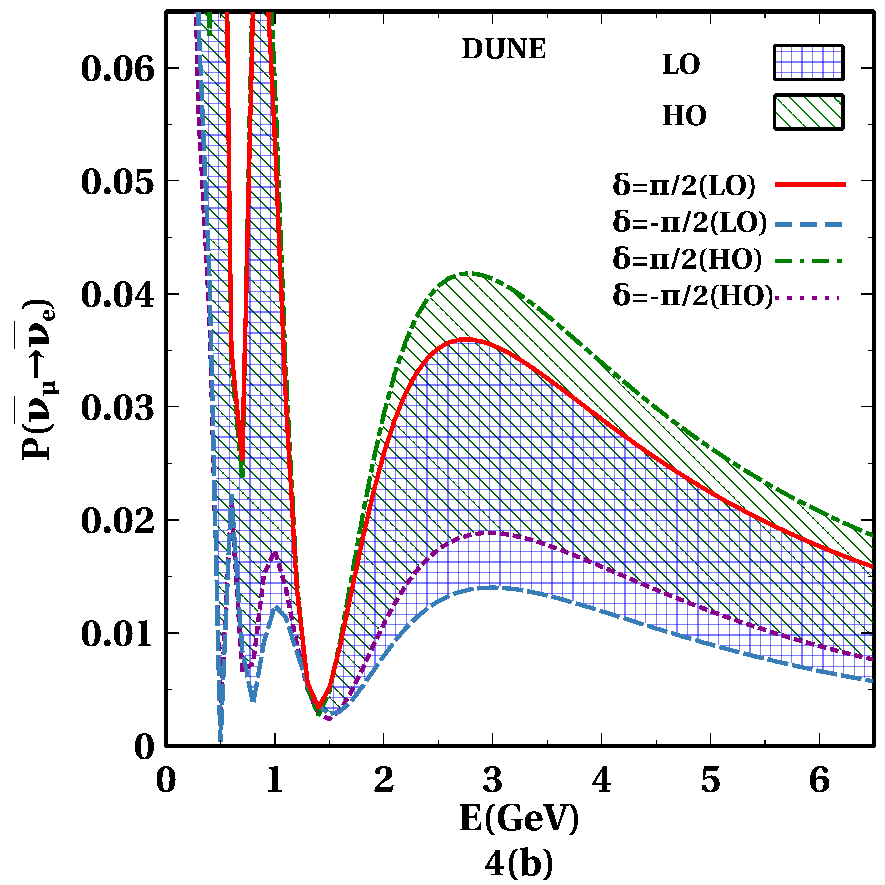, height=6cm, width=6cm}
\epsfig{file=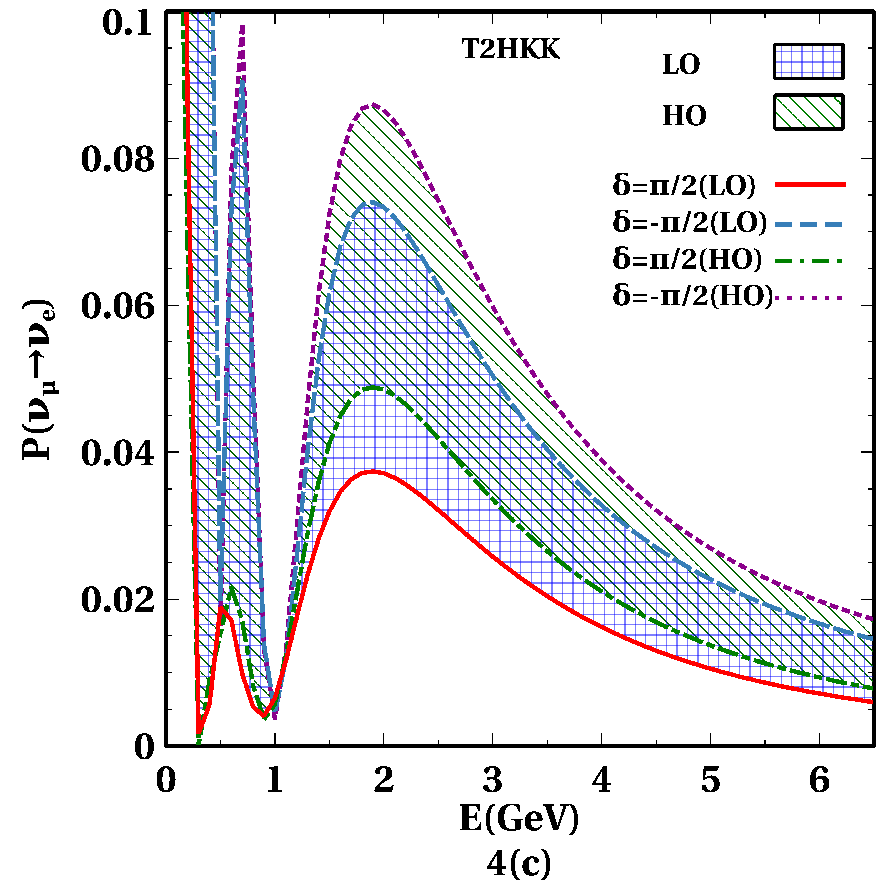, height=6cm, width=6cm}\hspace{1cm}
\epsfig{file=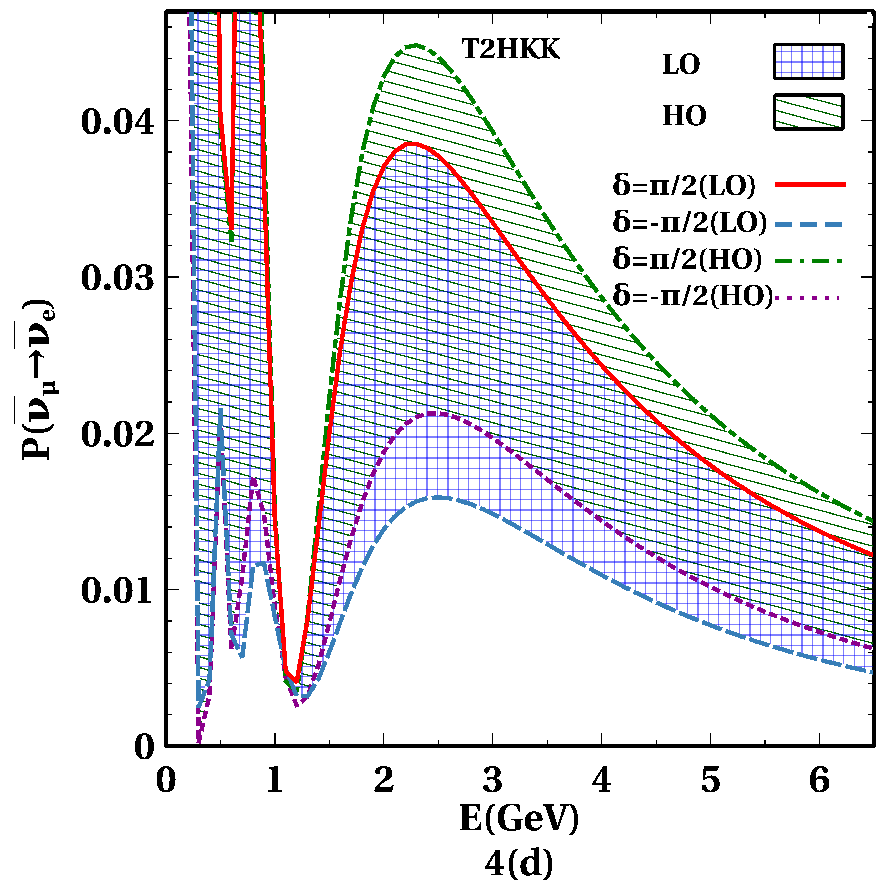, height=6cm, width=6cm}
\caption{~\label{fig4}The appearance probability for neutrino(antineutrino) mode as function of neutrino beam energy $E$ for DUNE(first row) and
T2HKK(second row) assuming NSI parameters $\epsilon_{e\mu}=0.04$ and $\delta_{e\mu}=[-\pi,\pi]$. The
octant degeneracy is represented for neutrino mode(left panel) and for antineutrino mode(right panel).
The band comes due to variation of $\delta,\delta_{e\mu}\in[-\pi,\pi]$ and the boundaries of the bands correspond to
$\delta=\pm\pi/2$ for LO and HO. The value of $\theta_{23}=42^{o}$ for LO and $\theta_{23}=48^{o}$ for HO.}
\end{figure}

 The oscillation probabilities for disappearance channels $1-P(\nu_{\mu}\rightarrow\nu_{e})$(neutrino) and
 $1-P(\bar{\nu_{\mu}}\rightarrow\bar{\nu_{e}})$(antineutrino) show
 main contribution from $\sin^{2}2\theta_{23}$. In case, $\theta_{23}$ is not maximal, we have two possibilities:
 either $\cos2\theta_{23}>0$ or $\cos2\theta_{23}<0$. This ambiguity creates two solutions $(\theta_{23},\delta)$ and $(90^{o}-\theta_{23},\delta')$.
 The resolution of uncertainty in $\theta_{23}$ due to octant degeneracy is important for precise measurement of $\theta_{23}$. We
 study this degeneracy in both neutrino and antineutrino mode with DUNE and T2HKK assuming NSI parameters
 $\epsilon_{e\mu}=0.04$ and $\delta_{e\mu}=[-\pi,\pi]$(Fig.(\ref{fig4})).

\begin{figure}[htp]
\centering
\epsfig{file=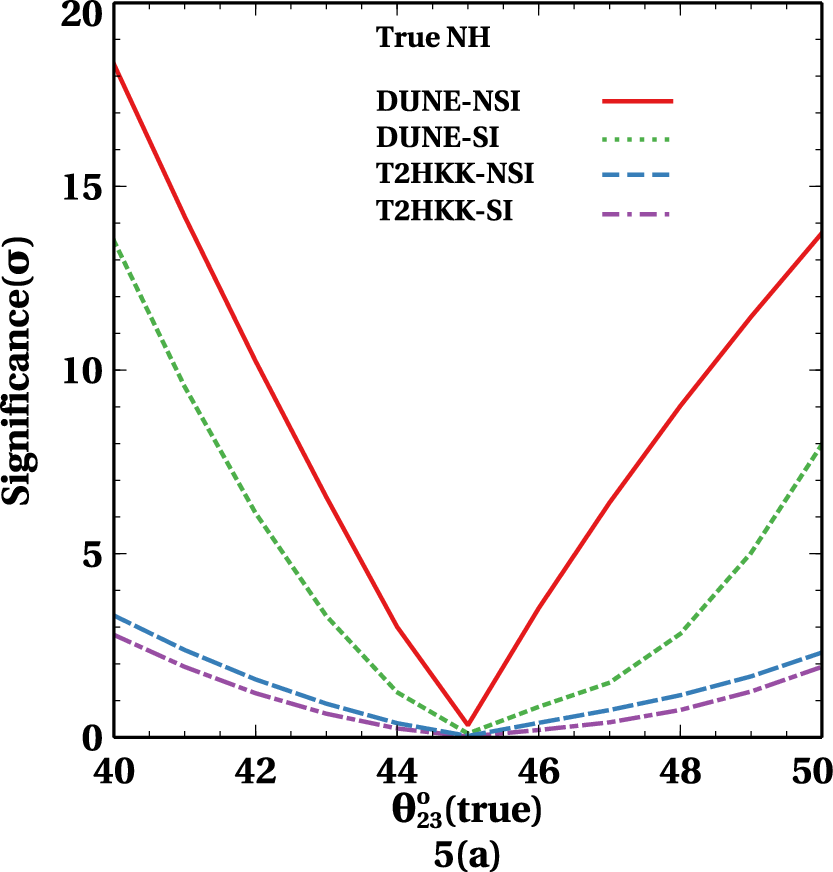, height=6cm, width=6cm}\hspace{1cm}
\epsfig{file=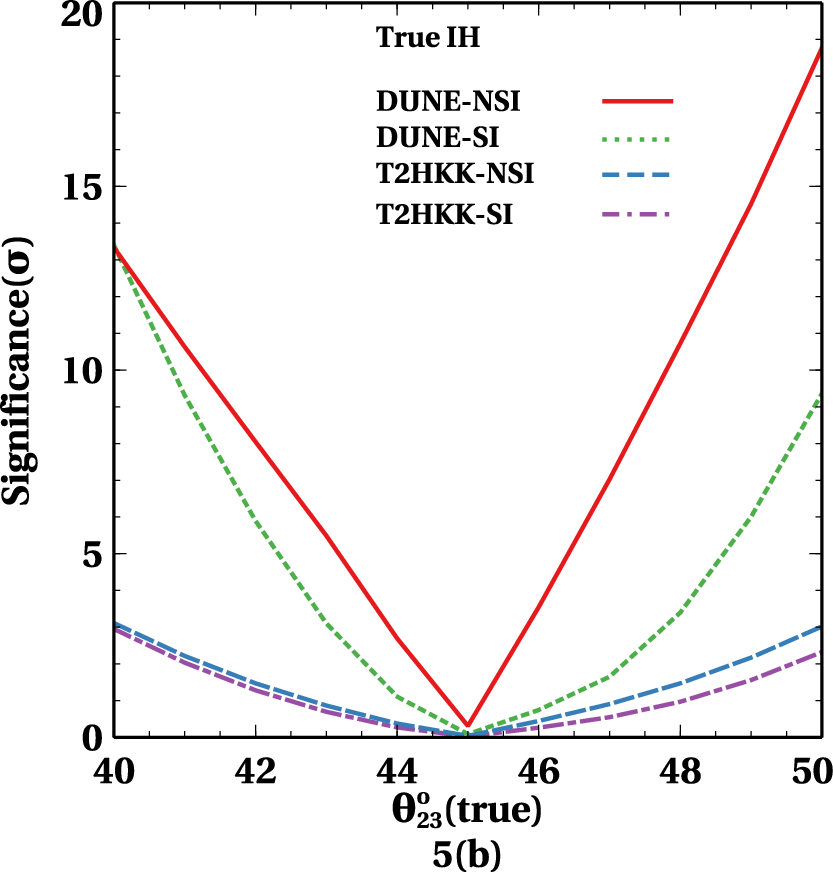, height=6cm, width=6cm}

\caption{~\label{fig5}Sensitivity plots for $\theta_{23}$ octant with true NH(left) and true IH(right), assuming NSI
parameters $\epsilon_{e\mu}=0.04$ and $\delta_{e\mu}=[-\pi,\pi]$.}
\end{figure}

To study the $\theta_{23}$ octant degeneracy, we have obtained octant sensitivity plot for true NH(true IH) in Fig.(\ref{fig5}).
 We plot significance($\sigma$) as a function of $\theta_{23}(true)$. One can define octant sensitivity by considering LO
 in true spectrum(HO in true spectrum) and HO in test spectrum(LO in test spectrum). The octant sensitivity tells us about the ability
 of an experiment to distinguish the $\theta_{23}$ lower octant from its higher octant. In order to obtain sensitivity plots for octant
 degeneracy, we have marginalized over sign($\Delta m_{31}^{2}$), $\delta$ and NSI parameter $\epsilon$ and the $\chi^{2}$ in true NH case is

\begin{eqnarray}
\nonumber
 \chi^{2}_{NH}&\equiv&\min_{(\delta,\theta_{23},\Delta m_{31}^{2},\epsilon)_{test}}\sum_{i=1}^{x}\sum_{j,k=1}^{2}\\ \nonumber
 &&\frac{\bigl[N_{NH}^{i,j,k}(\delta,\theta_{23},\Delta m_{31}^{2},\epsilon)_{true}-N_{NH}^{i,j,k}
               (\delta,\theta_{23},\Delta m_{31}^{2},\epsilon)_{test}\bigr]^{2}}
              {N_{NH}^{i,j,k}(\delta,\theta_{23},\Delta m_{31}^{2},\epsilon)_{true}},\\
\end{eqnarray}

\subsection{$CP$ phase degeneracy}
\begin{figure}[htp]
\centering
\epsfig{file=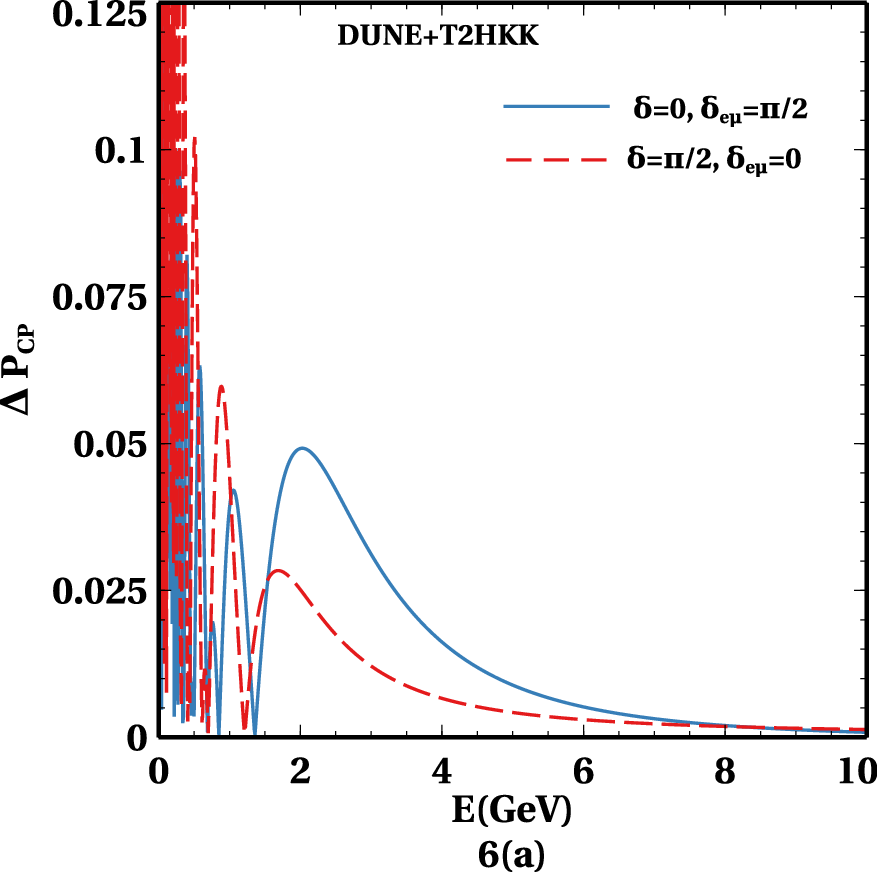, height=6cm, width=6cm}\hspace{1cm}
\epsfig{file=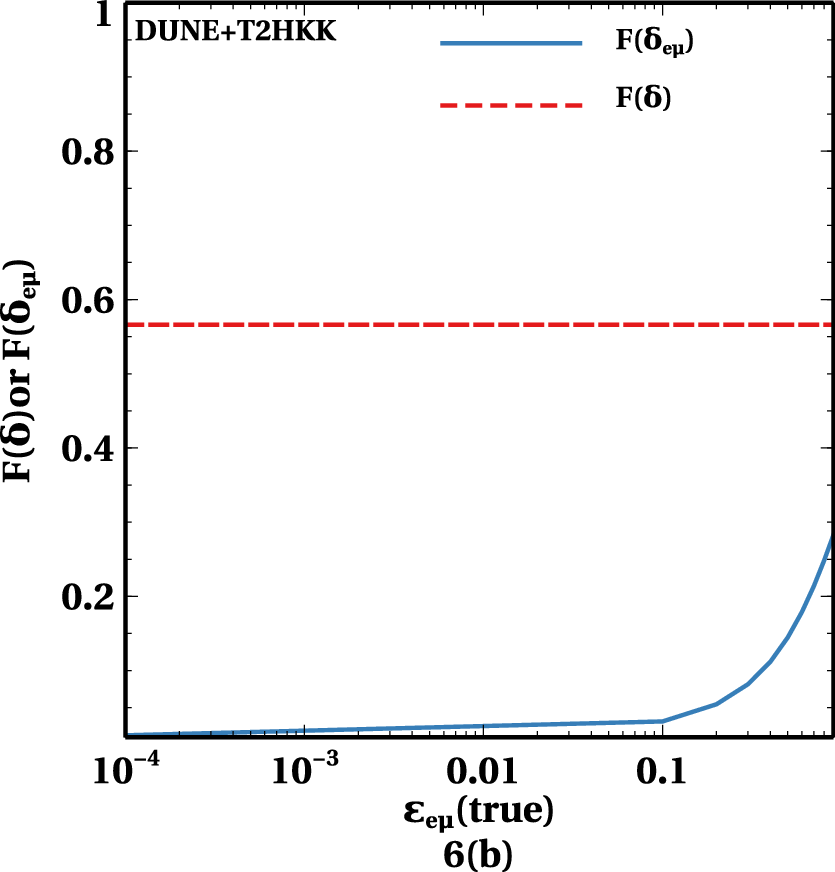, height=6cm, width=6cm}
\caption{~\label{fig3}The $CP$ asymmetry as function of neutrino beam energy $E$ for DUNE+T2HKK(left panel). The
$CP$ fractions corresponding to $\delta$ and $\delta_{e\mu}$ for which significance $\geq$ 3$\sigma$(right panel).}
\end{figure}

In presence of new physics, there may appear additional sources of $CP$ violation other than due to Dirac-type $CP$ phase $\delta$.
The $CP$ effects will, in general, include contributions from both standard and non-standard $CP$ phases.
As the value of $CP$ phases $\delta$ and $\delta_{e\mu}$ is not known and possible existence of $CP$ violation in nature, we 
look for all possible values of $\delta(true),\delta_{e\mu}(true)$ which are distinct from the $CP$ conserving values of $\delta$ and $\delta_{e\mu}$.
For this purpose, we define $CP$ fraction\cite{cpfraction} $F(\delta)(F(\delta_{e\mu}))$ as the fraction of total permitted range of $\delta(true)(\delta_{e\mu}(true))$ i.e. [$-\pi,\pi$] where $CP$ violation effects
corresponding to standard(non-standard) $CP$ phases can be explored. Also, we have excluded the $CP$ conserving values by marginalizing over $\delta(test)$($\delta_{e\mu}(test)$) for $\{0,\pi\}$ which implies that the
value of $\delta(test)$($\delta_{e\mu}(test)$) get fixed to its $CP$ conserving values $0$ or $\pi$(Fig.(\ref{fig3}(b))). So, we can write $CP$ fraction $F(\delta)$($F(\delta_{e\mu})$) at 3$\sigma$ C.L. $=\frac{\text{Total range of } \delta(true)(\delta_{e\mu}(true)) \text{ values above 3}\sigma \text{ C.L.}}{\text{Total permitted range of $\delta(true)(\delta_{e\mu}(true)) $}([-\pi,\pi])}$ which will be discussed in detail, in section 5. 

To resolve $CP$ phase degeneracy, we need to find out the $CP$ sensitivity with which an experiment can distinguish between $CP$ conserving cases
and $CP$ violating cases. In standard oscillations there is only one degree of freedom in $\chi^{2}$-function corresponding to standard $CP$
phase $\delta$. However, in case of neutrino oscillations with NSI there are two degrees of freedom due to standard and non-standard $CP$
phases $(\delta,\delta_{e\mu})$. We define the $\chi^{2}$-function as

\begin{equation}
 \chi^{2}\equiv \min_{\delta,\epsilon_{e\mu},\delta_{e\mu}}^{}\sum_{i=1}^{x}\sum_{j}^{2}\frac{[N_{true}^{i,j}(\delta,\epsilon_{e\mu},
 \delta_{e\mu})-N_{test}^{i,j}
 (\delta,\epsilon_{e\mu},\delta_{e\mu})]^2}{N_{true}^{i,j}(\delta,\epsilon_{e\mu},\delta_{e\mu})},
\end{equation}
where $N_{true}^{i,j}$ are the number of true events for $(\delta,\delta_{e\mu})$ in the range $[-\pi,\pi]$
and $N_{test}^{i,j}$ are the number of test events in $(i,j)^{th}$
bin for $(\delta,\delta_{e\mu})$ with $\{0,\pi\}$.

\begin{figure}[htp]
\centering
\epsfig{file=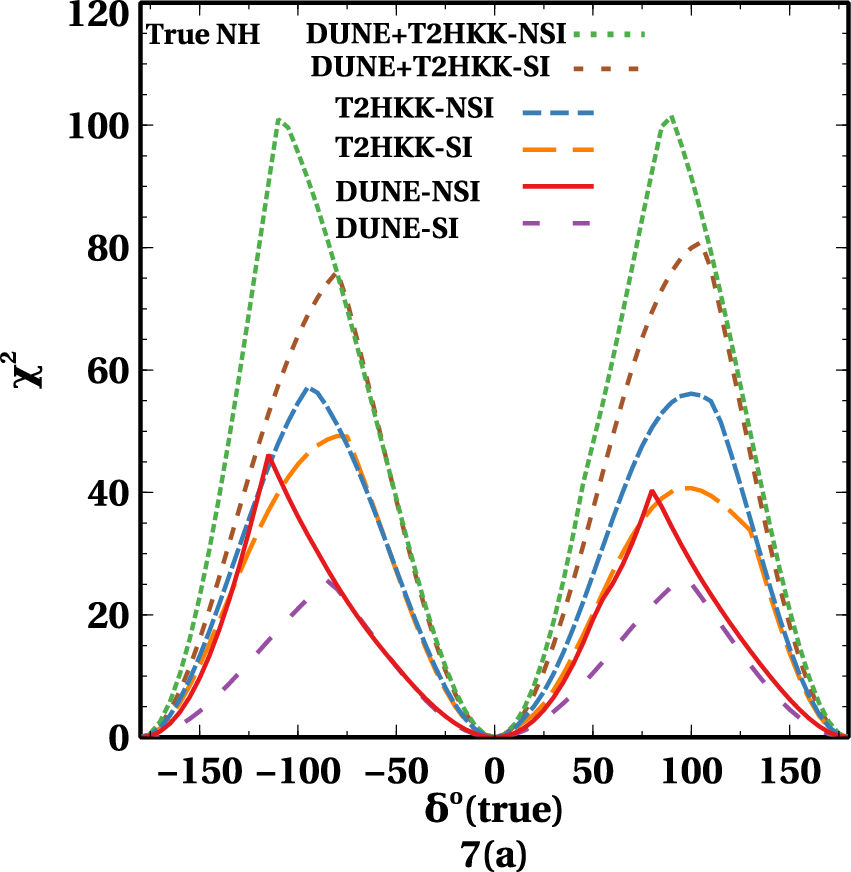, height=6cm, width=6cm}\hspace{1cm}
\epsfig{file=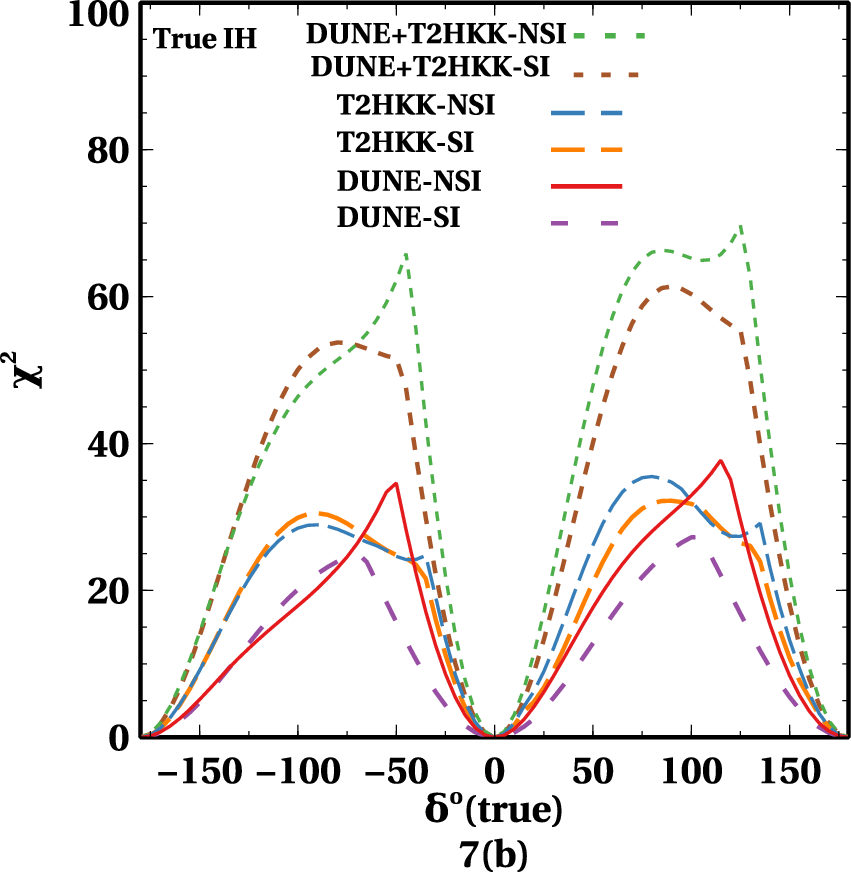, height=6cm, width=6cm}

\caption{~\label{fig7}Sensitivity plots for $CP$ violation with true NH(left) and true IH(right) assuming NSI parameters
$\epsilon_{e\mu}=0.04$ and $\delta_{e\mu}(true)=\pi/2$.}
\end{figure}

To study $CP$ violation discovery, we obtain plot for $\chi^{2}$ as a function of $\delta(true)$ for true NH and
true IH for DUNE, T2HKK and DUNE+T2HKK experiments(Fig.(\ref{fig7})).

\section{Results and Discussion}
The degeneracy in sign of $\Delta m_{31}^{2}$ and $\delta$ can be resolved with the experiments which involve matter effects
such as DUNE and T2HKK. In left(right) panel of Fig.(\ref{fig1}), we have shown the mass hierarchy degeneracy assuming
all($\epsilon_{e\mu}, \delta_{e\mu}$) NSI parameters, along with corresponding $CP$ phases, to be non-zero for DUNE and T2HKK experiments.
The meshed region between solid and dashed lines correspond to the normal hierarchy(NH)
whereas the forward-diagonal region between dash-dotted and dotted lines correspond to the inverted hierarchy(IH) of
neutrino masses. The normal and inverted
hierarchy regions overlap for DUNE and T2HKK experiments due to contributions from all the non-zero NSI parameters(left panel) making it
difficult to resolve the mass hierarchy degeneracy in these experiments. However, there is no overlap in the NH and IH regions
for DUNE and T2HKK for energy range 1 to 4 GeV assuming only $\epsilon_{e\mu}$ and corresponding phase to be non-zero(right panel).
Furthermore, the future long baseline experiments such as DUNE and T2HKK with higher statistics and better energy resolution, focusing on
neutrino beam energy between 1 to 4 GeV energy bracket, may provide better opportunity to resolve mass hierarchy
degeneracy(Figs.(\ref{fig1}(b) and \ref{fig1}(d))). However, for the same energy range DUNE has brighter prospects to resolve the mass degeneracy
than T2HKK.

In Fig.(\ref{fig1a}), we have obtained the sensitivity plots for mass hierarchy in presence of all($\epsilon_{e\mu},\delta_{e\mu}$)
NSI parameters for DUNE and T2HKK in first(second) row with true NH and true IH. Here, all NSI parameters referred to $\epsilon_{ee},\epsilon_{e\mu},
\epsilon_{e\tau},\epsilon_{\mu\mu},\epsilon_{\mu\tau},\epsilon_{\tau\tau}$ with corresponding $CP$ phases and their true and test values has been taken
from Table \ref{tab1}. For ready reference, we have, also, shown the sensitivity plots for the SI case in Fig.(\ref{fig1a}(e)) and Fig.(\ref{fig1a}(f)). We find from Fig.(\ref{fig1a}(a)) and Fig.(\ref{fig1a}(b)), that there is no distinction between true NH and true
IH sensitivities for any value of $\delta(true)$ at nearly $20\sigma$($3\sigma$) C.L. for DUNE(T2HKK) experiments.
The hierarchy sensitivity get enhanced when only one NSI parameter $\epsilon_{e\mu}$ and its corresponding $CP$ phase $\delta_{e\mu}$ is present
as shown in Fig.(\ref{fig1a}(c)) and Fig.(\ref{fig1a}(d)). The DUNE experiment in case of true NH shows stronger hierarchy sensitivity for
$-180^{o}<\delta<0^{o}$ as compared to true IH and is maximum at $\delta\approx-60^{o}$ ($15\sigma$ C.L.),
whereas, for $0^{o}<\delta<180^{o}$ the hierarchy sensitivity in true IH case is stronger and is maximum at $\delta\approx40^{o}$($11\sigma$ C.L.).
For T2HKK experiment the hierarchy sensitivity of true NH case stronger than true IH case for the
region $-90^{o}<\delta<90^{o}$ and  lies nearly at $5\sigma$ C.L.. It can be noted from Fig.(\ref{fig1a}(a)) and Fig.(\ref{fig1a}(b)) that $\chi^2$ decreases as $\theta_{23}$ changes from lower to higher octant which is in contradistinction to the case with one off-diagonal NSI(Fig.(\ref{fig1a}(c)) and Fig.(\ref{fig1a}(d)) and SI case(Fig.(\ref{fig1a}(e)) and Fig.(\ref{fig1a}(f))). Consequent to the presence of multiple NSI parameters and long baselines of DUNE and T2HKK, $\theta_{23}$ measurement will be severely affected by the NSI-modified matter effects due to degeneracies between SI and NSI parameters and between NSI parameters. Also, the mass hierarchy sensitivity will be decreased as compared to SI case due to cancellation effects induced by the off-diagonal NSI parameters. In particular, for all NSI parameters, at lower value of $\theta_{23}$ the cancellation effects will be relatively small as compared to SI matter effects which is reflected as increase in the $\chi^2$ value. However, cancellation effects will be appreciable for larger value of $\theta_{23}$ resulting in low $\chi^2$.

In Fig.(\ref{fig4}), we have shown octant degeneracy in DUNE and T2HKK with NSI parameters $\epsilon_{e\mu}=0.04$ and $\delta_{e\mu}=[-\pi,\pi]$.
It is evident from Fig.(\ref{fig4}) that octant degeneracy can be resolved with DUNE and T2HKK experiments using combination of neutrino and
antineutrino oscillation modes.
In Fig.(\ref{fig4}), the meshed region between solid and dashed lines represent the lower octant(LO)
and the forward-diagonal region between dash-dotted and dotted lines represent the higher octant(HO).
The left(right) panel in Fig.(\ref{fig4}) represents neutrino(antineutrino) mode of DUNE and T2HKK experiments. For LO($\theta_{23}<45^{o}$),
with $\delta=\pi/2$ and for HO($\theta_{23}>45^{o}$) with $\delta=-\pi/2$, both DUNE and T2HKK can resolve the octant degeneracies with neutrino
mode only. Moreover, for LO  with $\delta=-\pi/2$ and HO with $\delta=\pi/2$, octant degeneracies can be resolved with both DUNE and T2HKK with
antineutrino mode only. Thus, in general, the neutrino and antineutrino modes are exigent to resolve octant degeneracy in DUNE and T2HKK experiments
with matter NSI.  Also, we find that the neutrino(antineutrino) beam energy bracket of 1 to 4 GeV can, simultaneously, resolve the mass hierarchy and
octant degeneracies in DUNE and T2HKK.  

In Fig.(\ref{fig5}), we have obtained the sensitivity plots of $\theta_{23}$ octant with true NH and true IH for DUNE and T2HKK.
It can be seen from Fig.(\ref{fig5}), the octant degeneracy can be resolved for both DUNE and T2HKK experiments for true NH and true IH.
The DUNE experiment shows strong sensitivity for LO in true NH($18\sigma$ C.L.) and for HO in true IH($19\sigma$ C.L.). The T2HKK experiment
has weak sensitivity to resolve octant degeneracy in both true NH and true IH case. We have, also, checked that there is not much improvement 
in $\theta_{23}$ octant sensitivity(over DUNE case) if we take DUNE and T2HKK conjunctively. For comparison, we have, also, shown the SI sensitivity curves of $\theta_{23}$ octant with true NH and true IH for DUNE and T2HKK. 

The current and future neutrino oscillation experiments are diligently aiming at measuring neutrino mass hierarchy and $CP$ violating phase $\delta$. In presence of NSI (for example, assuming $\epsilon_{e\mu}$ and $\delta_{e\mu}$ non-zero) the situation becomes more complicated due to presence of additional sources of $CP$ violation. The nature may intromit $CP$ violation for wide range of $\delta$($\delta_{e\mu}$). 

In Fig.(\ref{fig3}(a)), we have shown $CP$
asymmetry with neutrino beam energy $E$ for both $\delta$ and $\delta_{e\mu}$ for DUNE+T2HKK. In Fig.(\ref{fig3}(a)), the solid (dashed) line
represents $\delta=0,\delta_{e\mu}=\pi/2(\delta=\pi/2,\delta_{e\mu}=0)$ case whereas in Fig.(\ref{fig3}(b)), the solid (dashed) line represents
the $CP$ fraction corresponding to the $\delta_{e\mu}(\delta)$. In Fig.(\ref{fig3}(b)), instead of focusing on measurement of $CP$ phases($\delta,\delta_{e\mu}$) we have obtained all possible $\delta(true)$ and $\delta_{e\mu}(true)$ values which are different from $CP$ conserving values of $\delta$ and $\delta_{e\mu}$ at 3$\sigma$ C.L. for DUNE+T2HKK. We have shown the effect of real NSI parameter($\epsilon_{e\mu}$) on the discovery reach of $CP$ violation due to $\delta,\delta_{e\mu}$ in Fig.(\ref{fig3}(b)).

For real NSI case($\epsilon_{e\mu}\in [10^{-4},1.0],\delta_{e\mu}=0$), we calculate $CP$ fraction $F(\delta)$, i.e. fraction of $\delta(true)\in [-\pi,\pi]$ which is distinguishable from its $CP$ conserving values, at 3$\sigma$ C.L..
The parameter space for which significance is less than 3$\sigma$ C.L. has been
excluded, thus, $\delta(test)$ has been fixed to its $CP$ conserving values $\{0,\pi\}$. The $\chi^{2}$ has been minimized for $\delta(true)$ value from $-\pi$
to $\pi$ and the parameter space for which significance $\geq 3\sigma$ is considered.
It is evident from Fig.(\ref{fig3}(b)) that real NSI parameter($\epsilon_{e\mu}$) has no effect on the discovery reach of $CP$ violation due to $\delta$($F(\delta)$), within the available bound on $\epsilon_{e\mu}$, at DUNE+T2HKK.

For complex NSI($\epsilon_{e\mu}\in[10^{-4},1.0], \delta_{e\mu}\neq0$), with $\delta=0$ i.e. $CP$ violation is only due to NSI phase $\delta_{e\mu}$, we have calculated $CP$ fraction $F(\delta_{e\mu})$ on similar lines as that for $F(\delta)$ in real NSI case(replace $\delta$ by $\delta_{e\mu}$). It can be seen from Fig.(\ref{fig3}(b)) that $F(\delta_{e\mu})\in[0,0.27]$ for $\epsilon_{e\mu}\in[10^{-4},1.0]$. However, for longer baseline $F(\delta_{e\mu})$ may be larger even for small value(s) of $\epsilon_{e\mu}$. In this case, if $CP$ violation is not observed at shorter baselines then the larger value of $F(\delta_{e\mu})$ imply $CP$ violation due to matter NSI.
Also, it is observed from Fig.(\ref{fig3}(b)) that $CP$ fraction $F(\delta)=0.57$ implying that there exist certain range(s) of $\delta(true)$ for which $CP$ violation is undetectable because significance is less than 3$\sigma$. For the case when $CP$ violation is due to both $\delta$, $\delta_{e\mu}$ and $\delta(true)$ lies below 3$\sigma$ significance, $CP$ violation may be observed for certain range(s) of NSI parameters $\epsilon_{e\mu}$ and $\delta_{e\mu}$. However, it will be difficult to disentangle the source of $CP$ violation i.e whether it is due to SI phase $\delta$ or NSI phase $\delta_{e\mu}$.

Fig.(\ref{fig7}) depicts the $CP$ violation sensitivity plots for DUNE, T2HKK and DUNE+T2HKK experiments with true NH
and true IH. The $CP$ violation sensitivity of T2HKK is stronger than DUNE in true NH irrespective of the value of $\delta(true)$(Fig.(\ref{fig7}(a))). For true IH, $CP$ violation sensitivity of T2HKK is stronger than DUNE except for $25^{o}$ range of $\delta(true)$ in $\delta(true)<0$ region and $\delta(true)>0$ region for which $CP$ violation sensitivity of DUNE is stronger than T2HKK. The $CP$ violation sensitivity of DUNE experiment is stronger in true NH than in true IH and is maximum at $\delta=-115^{o}$($\delta=115^{o}$),
in true NH(IH). In combined analysis DUNE+T2HKK, the $CP$ violation sensitivity increases to $10\sigma$ and $8.2\sigma$ for true NH and true IH, respectively. In Fig.(\ref{fig7}(b)), the dip in the T2HKK sensitivity is due to the hierarchy-$\delta$ degeneracy. The $\delta(true)$ values at the dip correspond to wrong hierarchy solutions. SI $CP$ violation sensitivity curves for DUNE, T2HKK and DUNE+T2HKK experiments with true NH and true IH have, also, been shown.

\section{Conclusions}
In conclusion, we have investigated the sensitivities of DUNE and T2HKK experiments to resolve mass hierarchy and octant degeneracies in presence
of matter NSI. We have, also, studied the $CP$ phase degeneracy due to standard and non-standard $CP$ phases for DUNE+T2HKK.
The results are in consonance with the earlier studies on DUNE.
We have analyzed the standard parameter degeneracies in presence of matter NSI for T2HKK experiment. We find that the mass
hierarchy degeneracy cannot be resolved in presence of all NSI parameters due to their large experimental uncertainties(Figs. \ref{fig1}(a) and \ref{fig1}(c)
). However, it can be resolved for neutrino beam energy range 1 to 4 GeV in case of one non-zero NSI parameter $\epsilon_{e\mu}$ and
corresponding NSI $CP$ phase $\delta_{e\mu}$ for DUNE and T2HKK experiments(Figs. \ref{fig1}(b) and \ref{fig1}(d)).

DUNE and T2HKK shows poor sensitivity to resolve mass hierarchy degeneracy in presence of all NSI parameters(Figs.(\ref{fig1a}(a)-\ref{fig1a}(b))).
However, the sensitivity to mass hierarchy get enhanced when only one NSI parameter $\epsilon_{e\mu}$ and its corresponding $CP$
phase $\delta_{e\mu}$ is present(Figs.(\ref{fig1a}(c)-\ref{fig1a}(d))). DUNE shows stronger hierarchy sensitivity for
$-180^{o}<\delta<0^{o}$  in true NH than true IH case with maximum sensitivity at $\delta\approx-60^{o}$($15\sigma$ C.L.),
whereas, for $0^{o}<\delta<180^{o}$ the hierarchy sensitivity in true IH case is stronger and is maximum at $\delta\approx40^{o}$($11\sigma$ C.L.).
 T2HKK shows stronger hierarchy sensitivity in true NH case than true IH case for the region $-90^{o}<\delta<90^{o}$ and lies nearly at $5\sigma$ C.L..

Furthermore, for LO($\theta_{23}<45^{o}$), with $\delta=\pi/2$ and for HO($\theta_{23}>45^{o}$) with $\delta=-\pi/2$, both DUNE and T2HKK
can resolve the octant degeneracies with neutrino mode only. Moreover, for LO  with $\delta=-\pi/2$ and HO with $\delta=\pi/2$, octant degeneracies
can be resolved with both DUNE and T2HKK with antineutrino mode only. Thus, combination of neutrino and antineutrino mode of DUNE and T2HKK can resolve
the octant degeneracy.

The octant degeneracy can be resolved for both DUNE and T2HKK experiments for true NH and true IH(Fig.(\ref{fig5})).
The DUNE experiment shows stronger sensitivity for LO in true NH($18\sigma$ C.L.) and for HO in true IH($19\sigma$ C.L.). The T2HKK experiment
has weak sensitivity to resolve octant degeneracy in both true NH and true IH case. We have, also, checked that there is not much improvement 
in $\theta_{23}$ octant sensitivity(over DUNE case) if we take DUNE and T2HKK conjunctively. 

The $CP$ asymmetry from non-standard $CP$ phase $\delta_{e\mu}$ is more than the standard $CP$ phase $\delta$ for neutrino beam energy 1.5 to 7 GeV(Fig.(\ref{fig3}(a))). From Fig.(\ref{fig3}(b)) we observe that real NSI parameter($\epsilon_{e\mu}$) has no effect on the discovery reach of $CP$ violation due to $\delta$($F(\delta)=0.57$), within the available bound on $\epsilon_{e\mu}$, at DUNE+T2HKK. Also, for complex NSI($\epsilon_{e\mu}\in[10^{-4},1], \delta_{e\mu}\neq0$) with $\delta=0$, $F(\delta_{e\mu})\in[0,0.27]$ for $\epsilon_{e\mu}\in[10^{-4},1]$. It increase with increase in $\epsilon_{e\mu}$ and is 0.27 when $\epsilon_{e\mu}=1$. Also, it is observed that $F(\delta)=0.57$ implying that there exist certain range(s) of $\delta(true)$ for which $CP$ violation is undetectable because significance is less than 3$\sigma$. For $\delta,\delta_{e\mu}\ne 0$, and $\delta(true)$ lies below 3$\sigma$ significance then $CP$ violation may be observed for certain range(s) of NSI parameters $\epsilon_{e\mu}$ and $\delta_{e\mu}$. However, it will be difficult to disentangle the source of $CP$ violation i.e whether it is due to SI phase $\delta$ or NSI phase $\delta_{e\mu}$.

The $CP$ violation sensitivity of T2HKK is stronger than DUNE in true NH(Fig.(\ref{fig7}(a))). However, there exist a $25^{o}$ range
of $\delta(true)$ in case of true IH for which $CP$ violation sensitivity of DUNE is stronger than T2HKK. In combined analysis DUNE+T2HKK, the $CP$ violation sensitivity increases to $10\sigma$ and $8.2\sigma$ for true NH and true IH, respectively.

\textbf{\Large{Data Availability}}\\
The experimental data used in the present analysis is taken from \cite{biggio,data} and is openly accessible.\\

\textbf{\Large{Conflicts of Interest}}\\
The authors declare that they have no conflicts of interest.

\textbf{\Large{Acknowledgements}}\\
The authors would like to thank Jogesh Rout for valuable discussions during this work. S. V. acknowledges the financial support provided
by University Grants Commission (UGC)-Basic Science Research(BSR), Government of India vide Grant No. F.20-2(03)/2013(BSR).
S. B. acknowledges the financial support provided by the Central University of Himachal Pradesh. The authors, also, acknowledge Department
of Physics and Astronomical Science for providing necessary facility to carry out this work.

\end{document}